\documentclass{revtex4}
\usepackage{amsmath,amsfonts}
\usepackage{epsfig,graphicx}
\usepackage{epstopdf}

\begin{document}

\title{Self-force of a scalar charge in the space-time of extreme charged anti-dilatonic wormhole}

\author{Arkady A. Popov\thanks{Email address: apopov@ksu.ru}}
\address{Institute of Mathematics and Mechanics, Kazan Federal University, 18 Kremlyovskaya St.,\\
Kazan, 420008, Russia\\
apopov@kpfu.ru}

\author{O. Aslan \thanks{Email address: alsucuk@gmail.com}}
\address{Institute of Mathematics and Mechanics, Kazan Federal University, 18 Kremlyovskaya St.,\\
Kazan, 420008, Russia\\
alsucuk@gmail.com}

\begin{abstract}
The self-interaction for a static scalar charge in the space-time of extreme charged anti-dilatonic wormhole is calculated.
We assume that the scalar charge is the source of massless scalar field with minimal coupling of the scalar field to the curvature of spacetime.
\end{abstract}


\maketitle
\section{Introduction}

Wormholes are topological handles in spacetime linking different universes or different
parts of the same universe. Interest in these configurations dates back to at least 1916 \cite{Flamm:1916} with revivals of activity
following the works of Einstein and Rosen in 1935 \cite{EinRos:1935} and the later series of
works initiated by Wheeler in 1955 \cite{Wheeler:1955}. A fresh interest in this topic
has been rekindled by the works of Morris and Thorne \cite{MorTho88} and of Morris, Thorne and Yurtsever \cite{MorThoYur88}.
In classical general relativity, it is well known that  traversable wormholes as solutions to the Einstein
equations can only exist with exotic matter which violates the null energy condition
$T{\mu \nu} u^\mu u^\nu \geq 0$ for any null vector field $u^\mu$.
Various models providing the wormhole existence include scalar fields \cite{BarVis:2000,SushKim:2002};
wormhole solutions in Einstein-Gauss-Bonnet theory \cite{RichSim:2007,KKK:2011};
wormholes geometries induced by quantum effects \cite{NOOO1:99,NOOO2:99};
wormhole solutions in semi-classical theory of gravity \cite{HochPopSush:97,Pop:2005,GarLobo:2007};
solutions in modified theories of gravity \cite{LoboOli:2009,GarLobo:2010,GarLobo:2011,DeBenHor:2012,CHKLO:2012};
modified teleparallel theories \cite{BoeHarLobo:2012} and their extensions \cite{Baha:2016}, etc.
The geometry of wormhole as well as a good introduction in the subject may be found
in the Visser book \cite{Visser:95} and in the review by Lobo \cite{Lobo:07}
Static spherically symmetric wormholes would look observationally
almost like black holes. One of the effects which can differ these two spacetimes is the self-interaction
for the charge. Self-interaction  effects  in  curved  spacetime  have  been  vigorously  explored;
for  an extensive  review,  see \cite{Detweiler:2005,Khusnutdinov:2005,CDAOW:2009,PoissonPoundVega:2011}.
The origin of this induced self-interaction resides on the nonlocal structure of the field caused
by the space-time curvature or nontrivial topology.
In flat space-time, the effect is determined by the derivative of acceleration of the charge.
For electrically charged particles in flat spacetime, the self-force is given
by the Abraham-Lorentz-Dirac formula\cite{Dirac:1938,Poisson:1999}.
In static curved space-times and space-times with nontrivial topology the self-force can be nonzero
even for the charge at rest (here and below the words ''at rest'' mean that the velocity of charge is collinear to
the timelike Killing vector which always exists in a static space-time).
Results of this type were obtained for a static particle in flat spacetimes of the topological defects\cite{Linet:1986,Linet2:1986,Smith:1990,Khus:1994,Khus:1995,Khus2:1995,LonMor:2002}.
The formal expression for the electromagnetic self-force in an arbitrary curved space-time was first
derived by DeWitt and Brehme \cite{DeW-Bre:1960} and a correction was later
provided by Hobbs \cite{Hobbs1:1968}. Mino, Sasaki, and Tanaka \cite{MST:1997} and independently Quinn and Wald
\cite{Quinn:1997} obtained similar expressions for the gravitational self-force on a point mass.
The self-force on a charge interacting with a massless minimally coupled scalar field was considered by Quinn \cite{Quinn:2000}.
More recently, a fresh interest in the topic has been  focused  on  the gravitational  self-force,  in  an  effort
to  model  the  inspiral  and  gravitational-wave emissions  of  a  binary  system  with  a  small  mass  ratio
\cite{BarackSago:2010,DienerVegaWardellDetweiler:2011,WarburtonAkcayBarackGairSago:2012}.
This interest has been prompted by the preparation of gravitational wave detectors which are capable of detecting
gravitational waves emitted when a compact object falls into a supermassive black hole.
For the Schwarzschild black hole the self-force on a static charge $q$ is repulsive, and it
has the dependence \cite{Vilenkin:79, SmithWill:80} $f \sim q^2/r^3$, where $r$ is the Schwarzschild
radial coordinate of the charge. Significant efforts were made to calculate the self-force on the background
of different types of the black holes \cite{DeWitt:64,MacGruder:78,Ori:95,Ori:97,Burko1:00,Burko2:00,
BarackOri:00,Barack:00,Lousto:00,BarackBurko:00,BurkoLiu:01,Nakano:01,Barack:01,Detweiler:01,
Barack1:02,PfenningPoisson:02,BarackLousto:02,BarackOri:02,BarackOri1:03,Mino:03,Det1:2003,BarackOri2:03,Det2:2003}.
The self-force for a charge at rest in static spherically symmetric wormhole spacetimes is determined
by the profile of wormhole throat and coupling of the field of charge with curvature of spacetime \cite{KhusBakh:2007,Linet:2008,Krasnikov:2008,BezKhu:09,Popov:10,Khu:10,Popov:13,Taylor:13,PopovAslan:15}.
It was shown \cite{BezKhu:09,Taylor:13} that there is an infinite set of values for the coupling constant of
the scalar field to the curvature of the spacetime for which the self-force on a static scalar charge diverges.
This set of values depends on the profile of the throat. The nature of this divergence is not entirely clear.

The aim of this paper is to analyze the self-force problem for a static pointlike scalar charge in an extreme
charged anti-dilatonic wormhole \cite{Clement,Popov2015}.
This problem has mathematical difficulties. The field of charge in the considered problem is determined
by the Green's function of some ordinary inhomogeneous differential equation of second order
(see equation (\ref{req}) below). This Green's function can be expressed in terms of independent solutions
of the corresponding homogeneous equation. As a rule, the Wronskian of these solutions coincides, up to
a normalization constant, with the coefficient in front of the delta-function in the Green's function equation.
In the case under consideration, this is not so. To solve this problem, we used the properties
of the delta-function (see (\ref{df})). This enables us to reduce the problem to the standard case.

The paper is organized as follows. In section II we describe the gravitational background. In section III
we obtained the unrenormalized expression for self potential of the static scalar charge on the considered
background. Section IV describes the procedure of renormalization of the self potential and the final result.

Throughout this paper, we use units $c=  G=  1$.

\maketitle

\section{Background}\label{Sec:Space-Time}
Let us choose the line element of space-time in the following form \cite{Clement,Popov2015}
\begin{equation} \label{metric}
ds^2 = -e^{-2\alpha(r)} dt^2 + e^{2\alpha(r)} \left[d r^2 + \left( r^2+Q^2 \right)\left( d\theta^2 +\sin^2 \theta d \varphi^2 \right)\right],
\end{equation}
where $-\infty < r < \infty, \theta \in [0, \pi], \varphi \in [0, 2\pi]$,
\begin{eqnarray}
2\alpha(r) &=&  \frac{\pi^2}{4}- \arctan^2 \left(\frac{r}{|{Q}|}\right).
\end{eqnarray}
This metric describes extreme charged anti-dilatonic wormhole, for which following relations are valid
\begin{eqnarray}
Q_{\phi}=M=\frac{\displaystyle \pi^2}{\displaystyle 2} |Q|,
\end{eqnarray}
where $Q$ - electric charge of the wormhole, $M$ - mass of the wormhole and $Q_{\phi}$ - dilatonic charge of the wormhole.

\section{Self potential of the static scalar charge}\label{Sec:Scalar}
Let us consider a scalar field $\phi$ with scalar source $j$. The corresponding
field equation has the form
\begin{equation} \label{meq}
{\phi}^{;\mu}_{;\mu}  = -4 \pi j = -4\pi q \int
\delta^{(4)}(x^\mu,\tilde x^\mu(\tau)) \frac{d\tau}{ \sqrt{-g^{(4)}}},
\end{equation}
where $g^{(4)}$ is the determinant of the metric $g_{\mu \nu}$, $q$ is the scalar charge and $\tau$ is its proper time. The world line of the charge is given by $\tilde x^\mu(\tau)$.
For a particle at rest in static space-time (\ref{metric}) field equation (\ref{meq}) can be rewritten as follows
\begin{eqnarray}
&& \left\{ \frac{}{} e^{-2\alpha}  \left[ \frac{\partial^2}{\partial r^2}
+ \frac{2r}{\left( r^2+Q^2 \right)} \frac{\partial}{\partial r}
 +\frac{1}{(r^2+Q^2)}\left(\frac{\partial^2}{\partial \theta^2}+\cot \theta\frac{\partial}{\partial \theta}
\right. \right. \right. \nonumber \\ && \left. \left. \left.
+\frac{1}{\sin^2(\theta)}\frac{\partial^2}{\partial \varphi^2} \right) \right]  \right\}
\phi(r, \theta, \varphi; \tilde r, \tilde \theta, \tilde \varphi)
=-\frac{4 \pi q \delta(r, \tilde r) \delta(\theta, \tilde \theta) \delta(\varphi, \tilde \varphi)}{e^{3\alpha} (r^2+Q^2) \sin \theta},
\end{eqnarray}
where we take into account $d\tau/dt=\sqrt{-g_{tt}}={e^{-\alpha(r)}}$ for the static charge.
Due to spherical symmetry of the problem under consideration we represent the potential in the following form
\begin{equation} \label{sum}
\phi = 4\pi q \sum_{l,m} Y_{lm}(\Omega)Y^*_{lm}(\tilde \Omega) g_l(r, \tilde r)= q \sum_{l=0}^\infty
\left(2l+1\right) P_l(\cos\gamma) g_l(r,\tilde r),
\end{equation}
where $Y_{lm}(\Omega)$ is the spherical functions of argument $\Omega = (\theta, \varphi)$, $\cos \gamma \equiv \cos \theta \cos \tilde \theta
+\sin \theta \sin \tilde \theta \cos(\varphi-\tilde \varphi)$. The radial part, $g_{l}$, satisfies the equation
\begin{equation} \label{req}
 \frac{\partial^2 g_l(r, \tilde r)}{\partial r^2}
+ \frac{2 r}{\left( r^2+Q^2 \right)} \frac{\partial g_l(r, \tilde r)}{\partial r} - \frac{l(l+1)}{(r^2+Q^2)}
g_l(r, \tilde r)=-\frac{ \delta(r, \tilde r) }{ e^{\alpha(r)}(r^2+Q^2)}.
\end{equation}
Introducing a new function
\begin{equation} \label{Gl}
G_l(r, \tilde r)=  e^{\alpha(\tilde r)} g_l(r, \tilde r)
\end{equation}
and taking into account
\begin{equation} \label{df}
 \frac{ \delta(r, \tilde r) }{ e^{\alpha(r)}} =
 \frac{ \delta(r, \tilde r) }{ e^{\alpha(\tilde r )}}
\end{equation}
one can obtain the following equation for $G_l(r, \tilde r)$
\begin{equation} \label{g}
 \frac{\partial^2 G_l(r, \tilde r)}{\partial r^2}
+ \frac{2 r}{\left( r^2+Q^2 \right)} \frac{\partial G_l(r, \tilde r)}{\partial r} -\frac{l(l+1)}{(r^2+Q^2)} G_l(r, \tilde r)
=-\frac{ \delta(r, \tilde r) }{(r^2+Q^2)}.
\end{equation}
Two independent solutions of corresponding homogeneous equation will be denoted by $\Psi_1(r)$ and $\Psi_2(r)$
\begin{equation} \label{Psi}
\frac{d^2 \Psi}{d r^2}+ \frac{2 r}{\left( r^2+Q^2 \right)} \frac{d \Psi}{d r} -\frac{l(l+1)}{(r^2+Q^2)} \Psi =0.
\end{equation}
$\Psi_1(r)$ is chosen to be the solution which is equal to zero at $r \to +\infty$ and divergent at $r \to -\infty$.
$\Psi_2(r)$ is chosen to be the solution which is equal to zero at $r \to -\infty$ and divergent at $r \to +\infty$.
That is,
\begin{eqnarray}\label{condlimit}
\lim_{r \to +\infty}\Psi_1 &=& 0,\ \lim_{r \to +\infty}\Psi_2 = \infty, \nonumber \\
 \lim_{r \to -\infty}\Psi_1 &=& \infty,\ \lim_{r \to -\infty}\Psi_2 =0.
\end{eqnarray}
Then one can represent the solution of equation (\ref{g}) in the following form
\begin{equation}\label{radialform}
G_l = \theta(r-\tilde r) \Psi_1(r) \Psi_2 (\tilde r) +
\theta(\tilde r-r) \Psi_1(\tilde r) \Psi_2 (r).
\end{equation}
Normalization of $\Psi$ is achieved by integrating (\ref{g}) with respect to $ r$ from $(\tilde r-\epsilon)$ to $(\tilde r+\epsilon)$
and letting $\epsilon \rightarrow 0$. This results in the Wronskian condition
\begin{equation}\label{wronskian1}
W(\Psi_1,\Psi_2) = \Psi_1 \frac{d \Psi_2}{d r} - \Psi_2 \frac{d \Psi_1}{d r}  =
\frac 1{r^2+Q^2}.
\end{equation}
We consider the radial equation (\ref{Psi}) in domains $r > 0$ and $r < 0$.  We may easily construct independent
solutions of this equation for these two domains separately
\begin{eqnarray} \label{expl}
\phi^1_+(r)&=& P_l\left(\frac{i r}{|Q|}\right), \quad \phi^2_+(r)= Q_l\left(\frac{i r}{|Q|}\right), \quad r > 0, \nonumber \\
\phi^1_-(r)&=& P_l\left(\frac{-i r}{|Q|}\right), \quad \phi^2_-(r)= Q_l\left(\frac{-i r}{|Q|}\right), \quad r < 0,
\end{eqnarray}
where $P_l$ and $Q_l$  are the Legendre polynomials of the first and second kind.
 Asymptotically
\begin{equation} \label{asymptotes}
\phi^1_\pm|_{r\to \pm\infty} \sim r^l, \quad \phi^2_\pm|_{r\to \pm\infty} \sim r^{-l-1}.
\end{equation}
The Wronskian of these solutions has the following form (see Ref. \cite{BatErdV1})
\begin{eqnarray}
W(\phi^1_\pm,\phi^2_\pm) = \frac{\pm i |Q|}{r^2+Q^2}.\label{wronskian3}
\end{eqnarray}
The solutions over all space can be written as follows
\begin{eqnarray}\label{solutions}
\Psi_1 &=&\left\{ \begin{array}{lc}
                  \alpha^1_+ \phi^1_+ + \beta^1_+ \phi^2_+, & r >0 \\
                  \alpha^1_- \phi^1_- + \beta^1_- \phi^2_-, &
                  r <0
                \end{array} \right., \nonumber \\
\Psi_2 &=&\left\{ \begin{array}{lc}
                  \alpha^2_+ \phi^1_+ + \beta^2_+ \phi^2_+, & r >0 \\
                  \alpha^2_- \phi^1_- + \beta^2_- \phi^2_-, &
                  r <0
                \end{array} \right. ,
\end{eqnarray}
where $\alpha^{1,2}_{\pm}, \beta^{1,2}_{\pm}$ are constants.
Applying the boundary conditions(\ref{condlimit}) in (\ref{solutions}) and using (\ref{asymptotes}) we get
\begin{equation}
\alpha^1_+=0, \quad \alpha^2_-=0.
\end{equation}
Then the solutions (\ref{solutions}) reduce to the following form
\begin{eqnarray} \label{Psi2}
\Psi_1 &=&\left\{ \begin{array}{lc}
                  \beta^1_+ \phi^2_+, & r >0 \\
                  \alpha^1_- \phi^1_- + \beta^1_- \phi^2_-, &
                  r <0
                \end{array} \right., \nonumber \\
\Psi_2 &=&\left\{ \begin{array}{lc}
                  \alpha^2_+ \phi^1_+ + \beta^2_+ \phi^2_+, & r >0 \\
                  \beta^2_- \phi^2_-, &
                  r <0
                \end{array} \right. .
\end{eqnarray}
Substituting these expressions into the Wronskian and taking into account (\ref{wronskian3}) we get
\begin{eqnarray}\label{wr2}
W(\Psi_1, \Psi_2) 
                  & = & \left\{ \begin{array}{lc}
                   - \alpha^2_+ \beta^1_+  \frac{\displaystyle i |Q|}{\displaystyle (r^2+Q^2)}, & r >0 \\
                   - \alpha^1_- \beta^2_-  \frac{\displaystyle i |Q|}{\displaystyle  (r^2+Q^2)}, & r <0 \\
                \end{array} \right.
\end{eqnarray}
The Wronskian condition (\ref{wronskian1}) implies the
constraints on the coefficients:
\begin{eqnarray} \label{cons}
\alpha^2_+ \beta^1_+ = \alpha^1_- \beta^2_-  &=& \frac {- 1}{i |Q|}= \frac {i}{ |Q|}.
\end{eqnarray}
The matching conditions of solutions $\Psi_{1}(r), \Psi_{2}(r)$ at $r=0$ are the following
\begin{eqnarray}
&&\lim_{r \rightarrow +0}\Psi_{1}(r) = \lim_{r \rightarrow -0}\Psi_{1}(r), \quad
\lim_{r \rightarrow +0}\frac{d \Psi_{1}(r)}{d r} = \lim_{r \rightarrow -0} \frac{d \Psi_{1}(r)}{d r}, \nonumber \\
&&\lim_{r \rightarrow +0}\Psi_{2}(r) = \lim_{r \rightarrow -0}\Psi_{2}(r), \quad
\lim_{r \rightarrow +0}\frac{d \Psi_{2}(r)}{d r} = \lim_{r \rightarrow -0} \frac{d \Psi_{2}(r)}{d r}.
\end{eqnarray}
Using these conditions one can obtain the following relations \cite{KhusBakh:2007}
\begin{equation} \label{bcons1}
\beta^1_+ \phi^2_+(0) = \alpha^1_- \phi^1_-(0) + \beta^1_- \phi^2_-(0), \quad
\alpha^2_+ \phi^1_+(0) + \beta^2_+ \phi^2_+(0) = \beta^2_- \phi^2_-(0),
\end{equation}
\begin{equation} \label{bcons2}
\left. \beta^1_+ \frac{d \phi^2_+}{d r}\right|_0 = \left. \alpha^1_- \frac{d \phi^1_-}{d r}\right|_0 + \left. \beta^1_- \frac{d \phi^2_-}{d r}\right|_0, \
\left. \alpha^2_+ \frac{d \phi^1_+}{d r}\right|_0 + \left. \beta^2_+ \frac{d \phi^2_+}{d r}\right|_0 = \left. \beta^2_- \frac{d \phi^2_-}{d r}\right|_0.
\end{equation}
These relations can be rewritten as follows
\begin{eqnarray} \label{c1}
&& \alpha^1_- = \beta^1_+
\left.\frac{W(\phi_+^2,\phi_-^2)}{W(\phi_-^1,\phi_-^2)}\right|_0,\
\beta^1_- = \beta^1_+ \left.
\frac{W(\phi_-^1,\phi_+^2)}{W(\phi_-^1,\phi_-^2)} \right|_0, \nonumber \\
&& \alpha^2_+ = - \beta^2_-
\left.\frac{W(\phi_+^2,\phi_-^2)}{W(\phi_+^1,\phi_+^2)}\right|_0, \
\beta^2_+ = +\beta^2_-
\left.\frac{W(\phi_+^1,\phi_-^2)}{W(\phi_+^1,\phi_+^2)}\right|_0.
\end{eqnarray}
Taking into account (\ref{expl}) and using the relations (see Ref. \cite{BatErdV1})
\begin{eqnarray}
P_l(0) &=& \frac{\sqrt{\pi}}{\Gamma (\frac{1}{2} - \frac{l}{2})\Gamma (1+\frac{l}{2})},\ \ P_l'(0) = -\frac{2\sqrt{\pi}}{\Gamma (\frac{1}{2} + \frac{l}{2})\Gamma
(-\frac l2)},\\
Q_l(0) &=& \frac{\sqrt{\pi}}2 e^{-i\frac\pi 2 (l+1)} \frac{\Gamma (\frac 12 + \frac l2)}{\Gamma (1+\frac l2)},\ \ Q_l'(0) = \sqrt{\pi} e^{-i\frac\pi 2
l} \frac{\Gamma (1 + \frac l2)}{\Gamma (\frac 12+\frac l2)}
\end{eqnarray}
the Wronskians in (\ref{c1}) can be calculated as:
\begin{eqnarray}
&&W(\phi_+^2,\phi_-^2)|_0=-\frac{\pi (-1)^{l}}{|Q|}, \quad W(\phi_-^1,\phi_+^2)|_0=\frac{i\pi}{|Q|}(-1)^l, \quad
W(\phi_-^1,\phi_-^2)|_0=-\frac{i}{|Q|}, \nonumber\\
&&W(\phi_+^1,\phi_+^2)|_0=\frac{i}{|Q|}, \quad W(\phi_+^1,\phi_-^2)|_0=-\frac{i(-1)^l}{|Q|}.
\end{eqnarray}
Then, we can rewrite the equations (\ref{c1}) as
\begin{eqnarray} \label{c3}
&&\alpha^1_- =-i\pi(-1)^l \beta^1_+, \quad \beta^1_- = -\pi(-1)^l \beta^1_+, \nonumber \\
&&\alpha^2_+ = i\pi(-1)^l \beta^2_-, \qquad \beta^2_+ =-(-1)^l \beta^2_-.
\end{eqnarray}
Using these relations  and (\ref{radialform},\ref{expl},\ref{Psi2},\ref{cons}) we obtain for $r>\tilde r>0$
\begin{eqnarray}\label{gg}
G_l(r,\tilde r) & = & \frac {i}{|Q|}\phi^2_+(r)\phi^1_+(\tilde r) +
\left.\frac {i}{|Q|}\frac{W(\phi^1_+,\phi^2_-)}{W(\phi^2_+,\phi^2_-)}
\right|_0 \phi^2_+(r)\phi^2_+(\tilde r),   \nonumber\\
& = & \frac {i}{|Q|}\phi^2_+(r)\phi^1_+(\tilde r) +
\frac 1{\pi |Q|}\phi^2_+(r)\phi^2_+(\tilde r) \nonumber\\
& = & \frac {i}{|Q|}Q_l(z)P_l(\tilde z) +
\frac 1{\pi |Q}|Q_l(z)Q_l(\tilde z)
\end{eqnarray}
 where $z=\frac{\displaystyle ir}{\displaystyle |Q|}$  and  $\tilde z=\frac{\displaystyle i\tilde r}{\displaystyle |Q|}$.

After substituting this expression into (\ref{Gl}) and then into (\ref{sum}), one can obtain,
 for the case $r>\tilde r>0$ and $\theta=\tilde \theta, \varphi=\tilde \varphi$
\begin{eqnarray}
\phi(r,\tilde r) = q \frac{e^{-\alpha(\tilde r)}}{|Q|}\sum_{l=0}^\infty (2l+1) \left\{
iP_l(\tilde z)Q_l(z) + \frac 1\pi Q_l(\tilde z)Q_l(z)\right\}.
\end{eqnarray}
To find $\displaystyle\sum_{l=0}^\infty (2l+1)P_l(\tilde z)Q_l(z)$  we use the Heine formula \cite{BatErdV1}
\begin {displaymath}\sum_{l=0}^\infty (2l+1)P_l(\tilde z)Q_l(z) = \frac 1{z - \tilde z}.\end {displaymath}
To find $\displaystyle\sum_{l=0}^\infty (2l+1)Q_l(\tilde z)Q_l(z)$ we use the integral
representation for the Legendre function of the second kind \cite{BatErdV1}
\begin{equation}
Q_l (z) = \frac 12 \int_{-1}^1 \frac{P_l(t)}{z-t}dt,
\end{equation}
and get
\begin{eqnarray}
\sum_{l=0}^\infty (2l+1)Q_l(i \tilde x)Q_l(ix) &=& -\frac{\arctan x - \arctan \tilde x}{x - \tilde x}.
\end{eqnarray}
Therefore for $r>\tilde r>0$ and $\theta=\tilde \theta, \varphi=\tilde \varphi$
\begin {eqnarray} \label{finpsi}
\phi(r,\tilde r) &=&q e^{-\alpha(\tilde r)}\left[ \frac 1{r-\tilde r} -
\frac{\arctan \frac{\displaystyle r}{\displaystyle |Q|} - \arctan \frac{\displaystyle \tilde r}{\displaystyle |Q|}}{\pi (r - \tilde r)}\right].
\end {eqnarray}

\section{Renormalization and result}

The procedure of the self-force evaluation requires the renormalization
of a scalar potential $\phi(x; \tilde x)$ which is diverged in the
limit $x \rightarrow \tilde x$ (see, for example, papers \cite{1Roth:2004,2Roth:2004}).
This renormalization is achieved by subtracting the DeWitt–Schwinger
counterterm $\phi_{\mbox{\tiny \sl DS}}(x; \tilde x)$ from $\phi(x; \tilde x)$ and then letting $x \rightarrow \tilde x$
\begin{equation} \label{ren}
\phi_{ren}(x)=\lim_{\tilde x \rightarrow x}\left( \phi(x; \tilde x)
-\phi_{\mbox{\tiny DS}}(x; \tilde x) \right).
\end{equation}
For a scalar charge at rest in static curved space-time the DeWitt–Schwinger counterterm $\phi_{\mbox{\tiny \sl DS}}(x; \tilde x)$, which must be subtracted, has the following form \cite{Popov2011}
        \begin{equation} \label{phiDS}
        \phi_{\mbox{\tiny DS}}(x^i; \tilde x^i)=
        q \left(\frac{1}{\sqrt{2 \sigma}}
        +\frac{\partial g_{t t}(\tilde x)}{ \partial \tilde x^i}
        \frac{\sigma^{{i}}}{4g_{t t}(\tilde x)\sqrt{2 \sigma}}
        \right),
        \end{equation}
where \cite{Synge,Popov:2007}
\begin{eqnarray}
{\sigma^i}&=&-\left(x^i-\tilde x^i\right)
        -\frac12 \Gamma^{i}_{{j}{k}}\left(x^j-{\tilde x^j}\right)\left(x^k-{\tilde x^k}\right)
        \nonumber \\ &&
        -\frac16 \left( \Gamma^{i}_{{j}{m}} \Gamma^{m}_{{k}{l}}
        +\frac{\partial \Gamma^{i}_{{j}{k}}}{\partial {\tilde x^l}}\right)
        \left(x^j-{\tilde x^j}\right)\left(x^k-{\tilde x^k}\right)\left(x^l-{\tilde x^l}\right)
        +O\left(\left(x-{\tilde x}\right)^4\right), \nonumber \\
\sigma &=& \frac{g_{i j }(\tilde x)}{2}  {\sigma^i} {\sigma^j},
\end{eqnarray}
$\Gamma^{i}_{{j}{k}}$ are the Christoffel symbols calculated at point $\tilde x$.
The DeWitt–Schwinger counterterm $\phi_{\mbox{\tiny \sl DS}}(x; \tilde x)$ in the limit $\theta= \tilde \theta, \varphi=\tilde \varphi$ can be easily calculated using the metric (\ref{metric})
        \begin{equation} \label{phiDSf}
        \phi_{\mbox{\tiny DS}}(x^i; \tilde x^i)=\frac{q e^{-\alpha(\tilde r)}}{|r-\tilde r|}
        \end{equation}
Using the expression (\ref{finpsi}) for $\phi(\tilde r, r)$  we obtain
the renormalized expression for $\phi$ in domain $r > 0$
\begin{equation} \label{fin}
\phi_{ren}(r)=\lim_{\tilde r \rightarrow r}\left[ \phi(r, \tilde r) - \phi_{\mbox{\tiny DS}}(r, \tilde r) \right]=-\frac{q |Q| e^{-\alpha(r)}}{\pi (r^2+Q^2)}.
\end{equation}
$\phi_{ren}$ in domain $r < 0$ coincides with this expression because of the symmetry $r \leftrightarrow - r$ of the problem.
The limiting case $r \rightarrow \infty$ gives us the following result
\begin{equation}
\phi_{ren}(r)\approx-\frac{q |Q|}{\pi r^2}.
\end{equation}
The only nonzero component of the self-force is
\begin{equation} \label{sf}
F^r(r)=-\frac{q}{2} g^{rr} \frac{ \partial \phi_{ren}(r)}{ \partial r}
=-\frac{q^2 e^{-3 \alpha(r)} \left( \frac{\displaystyle r}{\displaystyle |Q|} - \frac{\displaystyle 1}{\displaystyle 2} \arctan(\frac{\displaystyle r}{\displaystyle |Q|}) \right)}{\pi Q^2 \left(1+ r^2/Q^2 \right)^2}
\end{equation}
\begin{center}
\includegraphics[width=5cm]{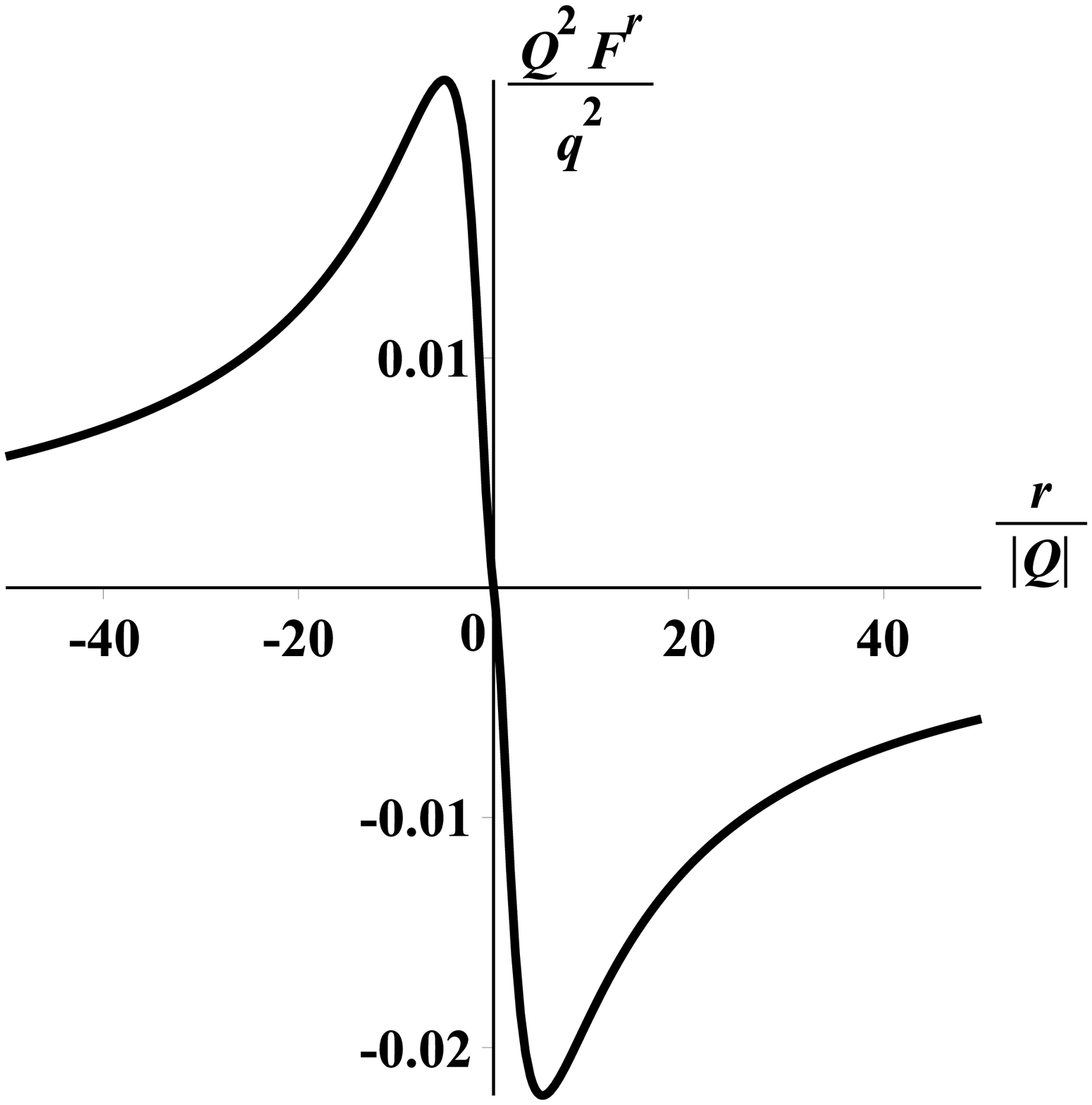}
\end{center}
Thus, we  have   obtained  an   analytic  expression (\ref{sf}) for  the self-force  on a  static  scalar  charge
in an extreme charged anti-dilatonic wormhole spacetime (\ref{metric}).

\section*{Acknowledgments}

The work is performed according to the Russian Government Program of Competitive Growth of Kazan Federal University.


\end{document}